\documentclass[aps,prl,twocolumn,groupedaddress]{revtex4}

\usepackage{graphicx} 
\usepackage{dcolumn}
\usepackage{bm}

\begin{document}
\title{Evidence for very strong electron-phonon coupling in 
YBa$_{2}$Cu$_{3}$O$_{6}$} 
\author{Guo-meng Zhao$^{*}$} 
\address{Department of Physics and Astronomy, California State 
University, Los Angeles, CA 90032, USA}

\begin{abstract}
 From the observed oxygen-isotope shift of the mid-infrared 
two-magnon 
absorption
peak of YBa$_{2}$Cu$_{3}$O$_{6}$, we evaluate the oxygen-isotope
effect on the in-plane antiferromagnetic exchange energy $J$. The
exchange energy $J$ in YBa$_{2}$Cu$_{3}$O$_{6}$ is found to decrease 
by
about 0.9$\%$ upon replacing $^{16}$O by $^{18}$O, which is slightly
larger than that (0.6$\%$) in La$_{2}$CuO$_{4}$.  From the oxygen-isotope
effects, we determine the lower limit of the polaron binding energy, which is 
about 1.7 eV for YBa$_{2}$Cu$_{3}$O$_{6}$ and 1.5 eV for 
La$_{2}$CuO$_{4}$, in quantitative agreement with 
angle-resolved photoemission data, optical conductivity data, and the 
parameter-free theoretical estimate.  The large polaron binding 
energies in the insulating parent compounds suggest that 
electron-phonon coupling should also be strong in doped 
superconducting cuprates and may play an essential role in 
high-temperature superconductivity.

\end{abstract}
\maketitle 
The behavior of undoped insulating compounds such as La$_{2}$CuO$_{4}$
and YBa$_{2}$Cu$_{3}$O$_{6}$ is of interest as the starting point for
discussion of the physics of cuprates, particularly with regards to
the microscopic pairing mechanism of
high-temperature superconductivity in doped systems. The
antiferromagnetic (AF) ordering found in the parent compounds
\cite{Vak,Tranq1,Tranq2,Brewer} signals a
strong electron-electron Coulomb correlation. On the other hand, 
there is overwhelming evidence that
electron-phonon coupling is very strong
in the cuprate superconductors
\cite{ZhaoAF,ZhaoYBCO,ZhaoLSCO,ZhaoNature97,ZhaoJPCM,McQueeney,Lanzara,SG,Mis,HoferPRL,Zhaoreview1,Zhaoreview2,Zhaoisotope,Lanzara01,Keller1,Zhou04}. 
In particular, various unconventional oxygen-isotope
effects Zhao and his
coworkers have observed since 1994 clearly indicate that the
electron-phonon interactions are so strong that polarons/bipolarons
are formed in doped cuprates
\cite{ZhaoAF,ZhaoYBCO,ZhaoLSCO,ZhaoNature97,ZhaoJPCM,Lanzara,SG,HoferPRL,Zhaoreview1,Zhaoreview2,Zhaoisotope,Keller1} and manganites
\cite{ZhaoNature96,Zhaobook},
in agreement with a theory of
high-temperature superconductivity \cite{ale} and the original
motivation for the discovery of high-temperature superconductivity
\cite{KAM86}.

Now it is well accepted that the parent compounds are
charge-transfer insulators and can be described by a three-band
Hubbard model. Recently, Eremin {\em et al.} \cite{Eremin} have 
considered strong
electron-phonon
coupling within the three-band Hubbard model.  They show that the 
antiferromagnetic exchange energy $J$ depends on the polaron
binding energy $E_{p}^{O}$ due to oxygen vibrations, on the polaron
binding energy $E_{p}^{Cu}$ due to copper vibrations, and on their
respective vibration frequencies $\omega_{O}$ and $\omega_{Cu}$. At
low temperatures, $J$ is given by \cite{Eremin}
\begin{equation}
J = J_{\circ}(1 +
\frac{3E_{p}^{O}\hbar\omega_{O}}{\Delta_{pd}^{2}}+\frac{3E_{p}^{Cu}\hbar\omega_{Cu}}{\Delta_{pd}^{2}}),
\end{equation}
where $\Delta_{pd}$ is the unrenormalized charge-transfer gap.  The 
bare superexchange interaction $J_{\circ}$ is obtained  within the fourth-order 
perturbation theory and its renormalization due to electron-phonon 
coupling takes place in the sixth-order term \cite{Eremin}.  That is why the 
effective exchange energy $J$ is not strongly renormalized by 
electron-phonon coupling, in contrast to a strong renormalization of 
the hopping integral by electron-phonon coupling \cite{Eremin}.

Since the polaron binding energy is independent of the masses of nuclear 
ions \cite{Alex2}, it is apparent from Eq.~1 that there should be an 
observable oxygen-isotope effect on $J$ if the polaron binding energy 
is comparable with $\Delta_{pd}$.  An increase of the oxygen mass 
leads to a decrease of the phonon energy, which in turn results in a 
reduction of the exchange energy $J$ according to Eq.~1.  
Quantitatively the oxygen-isotope effect on $J$ can be readily deduced 
from Eq.~1:
\begin{equation}\label{Ie8}
\frac{\Delta J}{J} =
(\frac{3E_{p}^{O}\hbar\omega_{O}}{\Delta_{pd}^{2}})(\frac{\Delta\omega_{O}}{\omega_{O}}).
\end{equation}

Zhao and his co-workers initiated studies of the
oxygen isotope effect on the AF ordering temperature in several
parent compounds \cite{ZhaoAF}. A noticeable oxygen-isotope shift of
$T_{N}$ (about 1.9 K) was
consistently observed in undoped La$_{2}$CuO$_{4}$ with $T_{N}$ = 315 
K
(Ref.~\cite{ZhaoAF}). From the observed oxygen isotope shift of 
$T_{N}$,
they found \cite{Zhaoreview1} that the antiferromagnetic exchange 
energy is reduced by
about 0.6$\%$ upon replacing $^{16}$O by $^{18}$O, i.e., $\Delta
J/J$ = $-$0.6$\%$.  However, this novel
isotope effect is negligible in electron-doped cuprates that do not
have apical oxygen \cite{ZhaoAF}. This implies that the apical
oxygen,  which can stabilize the $Q_{1}$-type Jahn-Teller
distortion,  enhances the electron-phonon interaction significantly.

Now a question arises: Does this isotope effect also exist in other
parent cuprates with apical oxygen? Recent mid-infrared spectra of 
YBa$_{2}$Cu$_{3}$O$_{6}$ crystals show that the two-magnon absorption 
peak is shifted down by 3.5 meV upon replacing $^{16}$O by $^{18}$O 
(Ref.~\cite{Grun}).  This isotope shift was explained in terms of the 
shift in the frequency of a high-energy oxygen vibration mode that 
assists the two-magnon absorption process~\cite{Grun}.  However, in order to 
reproduce the experimentally observed high-frequency spectral weight 
within this scenario, one requires a very large coupling constant that is  
one order of magnitude larger than an expected value \cite{Grun}.  Moreover, 
the deduced exchange energy (99.5 meV) \cite{Grun} is significantly 
lower than those (110-118 meV) inferred from neutron scattering and 
Raman scattering data \cite{Hayden,Blum} (see below).

Here we demonstrate 
that the phonon modes that assist the two-magnon absorption process are not the 
high-energy oxygen-related vibration modes, but the low-energy phonon 
modes (20-30 meV) which are mainly associated with copper vibrations 
and very strongly coupled to electrons \cite{Zhaoisotope,Gonnelli}.  
Within this scenario, the observed oxygen-isotope shift of the two-magnon 
absorption peak in YBa$_{2}$Cu$_{3}$O$_{6}$ is actually consistent with 
a significant 
oxygen-isotope effect on $J$, that is, $J$ decreases by about 0.9$\%$ upon 
replacing $^{16}$O by $^{18}$O, which is slightly
larger than that (0.6$\%$) in La$_{2}$CuO$_{4}$.  From the oxygen-isotope
effects, we determine the lower limit of the polaron binding energy, which is 
about 1.7 eV for YBa$_{2}$Cu$_{3}$O$_{6}$ and 1.5 eV for 
La$_{2}$CuO$_{4}$, in quantitative agreement with 
angle-resolved photoemission data, optical conductivity data, and the 
parameter-free theoretical estimate.  Such quantitative agreement strongly supports the model where the 
low-energy phonon modes mainly 
assist the two-magnon absorption process. The large 
polaron binding energies in the insulating parent compounds suggest 
that electron-phonon coupling should also be strong in doped 
superconducting cuprates and may play an essential role in 
high-temperature superconductivity.

It is known that Raman spectra can probe two-magnon scattering in
antiferromagnets. In contrast, two-magnon absorption in infrared (IR)
spectra is expected to be inactive. However, the excitation becomes
IR active when a phonon is simultaneously created \cite{Lorenz}. A 
photon with an
energy of $\hbar\omega_{IR} = 2.73J + \hbar\omega_{ph}$ is required
for this process in single-layer systems \cite{Lorenz,Choi}, where 
$\hbar\omega_{ph}$ is the phonon energy.  For double-layer systems such 
as YBa$_{2}$Cu$_{3}$O$_{6}$, the two-magnon absorption peak shifts up 
by about 1.6$J_{\perp}$ (Ref.~\cite{Grun}), that is, 
\begin{equation}\label{IR}
\hbar\omega_{IR} = 2.73J + 1.6J_{\perp}+\hbar\omega_{ph},
\end{equation}
where $J_{\perp}$ is the interlayer exchange energy within the 
bilayers.

From Eq.~\ref{IR}, we can determine the value of $J$ from the energy position 
of the two-magnon absorption peak if we are able to independently 
determine the phonon energy.  The phonon energy can be determined by 
the temperature dependence of the linewidth of the peak.  Choi {\em et 
al.} \cite{Choi} have shown that the temperature dependence of the 
linewidth of the absorption peak in Sr$_{2}$CuO$_{2}$Cl$_{2}$ is in 
quantitative agreement with the two-magnon absorption process that 
involves scattering by phonon modes centered at about 25 meV.  This is 
consistent with inelastic neutron scattering, which shows a broad 
maximum at about 27 meV in the phonon density of states of 
single-layer La$_{2-x}$Sr$_{x}$CuO$_{4}$ \cite{Mcq,AraiLSCO}.  
This is also in quantitative agreement with tunneling spectra in both 
optimally doped Bi$_{2}$Sr$_{2}$CaCu$_{2}$O$_{8+\delta}$ and 
YBa$_{2}$Cu$_{3}$O$_{7-\delta}$, which indicate a very large 
electron-phonon coupling constant (about 2.6) for the 20 meV phonon 
modes \cite{Zhaoisotope}. Substituting 
$\hbar\omega_{IR}$ = 363 meV (Ref.~\cite{Choi}) and $\hbar\omega_{ph}$ 
= 25 meV into Eq.~\ref{IR}, we find that $J$ = 124 meV for 
Sr$_{2}$CuO$_{2}$Cl$_{2}$.  The deduced $J$ = 124 meV from the IR 
spectrum is in perfect agreement with that (125 meV) evaluated 
precisely from the measured temperature dependence of the 
spin-correlation length \cite{Greven}.  This quantitative agreement 
provides additional evidence that the 25 meV phonon modes contribute 
to the phonon-assisted two-magnon excitation in this single-layer 
compound.

For another single-layer compound La$_{2}$CuO$_{4}$, the two-magnon 
absorption peak is at 410 meV (Ref.~\cite{Perkins}).  Substituting 
$\hbar\omega_{IR}$ = 410 meV and $\hbar\omega_{ph}$ = 25 meV into 
Eq.~\ref{IR} yields $J$ = 141 meV, in quantitative agreement with that 
(138$\pm$4 meV) deduced from the measured temperature dependence of 
the spin-correlation length \cite{Johnston}.  From the measured 
long-wave spin velocity ($\hbar c$ = 0.85$\pm$0.03 eV\AA) 
\cite{Aeppeli} and using a renormalization factor $Z_{c}$ = 1.14 
(Ref.~\cite{Jun}), we obtain $J$ = 139$\pm$5 meV, in excellent 
agreement with the above values.  Therefore, the average energy of the 
phonon modes assisting the two-magnon absorption process is also about 
25 meV in La$_{2}$CuO$_{4}$.

We can also extract the energy of the phonon modes assisting the two-magnon 
absorption process if we can reliably obtain the $J$ value from other 
independent experiments.  Two-magnon Raman scattering and resonant 
two-magnon Raman scattering experiments \cite{Blum} can independently determine 
the $J$ value.  For two-magnon scattering, the peak position 
$\hbar\omega_{R}$ is related to $J$ and $J_{\perp}$ as \cite{Blum}
\begin{equation}\label{R}
\hbar\omega_{R} = 2.8J + J_{\perp}.
\end{equation}
For resonant two-magnon Raman scattering, the first resonance peak 
occurs at \cite{Blum}
\begin{equation}\label{RR}
\hbar\omega^{1}_{res} = \Delta_{pd} + 2.9J.
\end{equation}
 
The Raman spectrum of 
Sr$_{2}$CuO$_{2}$Cl$_{2}$ shows a two-magnon scattering peak at 
$\hbar\omega_{R}$ = 355 meV (Ref.\cite{Tokura}).  Substituting 
$\hbar\omega_{R}$ = 355 meV and $J_{\perp}$ = 0 into Eq.~\ref{R} 
yields $J$ = 126.8 meV.  The measured temperature dependence of the 
spin-correlation length indicates $J$ = 125 meV (Ref.~\cite{Greven}).  
Hence, the $J$ value determined from the two independent experiments 
is 126$\pm$1 meV. Substituting $\hbar\omega_{IR}$ = 363 meV and $J$ = 126 meV into Eq.~\ref{IR} 
yields $\hbar\omega_{ph}$ = 19 meV, slightly lower than that (25 meV) extracted 
from the temperature dependence of linewidth of the absorption peak.

From Eqs.~\ref{R} and \ref{RR}, we can extract the exchange 
energy $J$ for YBa$_{2}$Cu$_{3}$O$_{6}$.  Substituting $J_{\perp}$ = 
11 meV (Ref.~\cite{Hayden}) and $\hbar\omega_{R}$ = 
342.7 meV (Ref.~\cite{Blum}) into Eq.~\ref{R} yields $J$ = 118 meV. Resonant two-magnon Raman scattering \cite{Blum} shows a resonance 
peak at $\Delta_{pd} + 328$ meV, that is, 
$\hbar\omega^{1}_{res} - \Delta_{pd}$ = 328 meV.  Then from 
Eq.~\ref{RR}, we find $J$ = 113 meV.  Neutron data 
\cite{Hayden} imply that $Z_{c}J$ = 125$\pm$5 meV, leading to $J$ = 
110$\pm$5 meV with $Z_{c}$ = 1.14 (Ref.~\cite{Jun}).  Thus, the $J$ 
value deduced from the three independent experiments is 114$\pm$4 meV.

\begin{figure}[htb]
    \includegraphics[height=6.2cm]{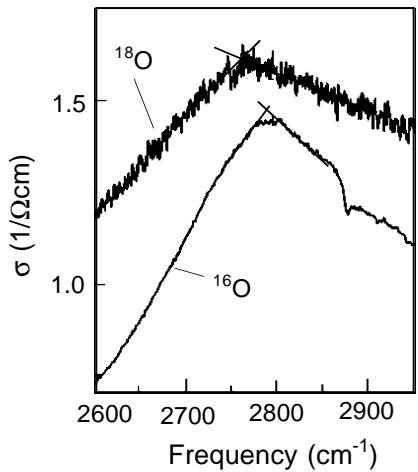}
 \caption[~]{The mid-infrared optical conductivity for the 
$^{16}$O and $^{18}$O samples of YBa$_{2}$Cu$_{3}$O$_{6}$ at $T$ = 4 
K.  The figure is reproduced from Ref.~\cite{Grun}.  Upon replacing 
$^{16}$O by $^{18}$O, the peak position shifts down by about 26.8 
cm$^{-1}$.}
\end{figure}

Now we analyze the observed oxygen-isotope shift of the two-magnon absorption 
peak for YBa$_{2}$Cu$_{3}$O$_{6}$ \cite{Grun}.  Fig.~1 shows the mid-infrared optical conductivity for 
the $^{16}$O and $^{18}$O samples of YBa$_{2}$Cu$_{3}$O$_{6}$ at $T$ = 
4 K.  The figure is reproduced from Ref.~\cite{Grun}.  The 
intersection point of the two straight lines defines the peak position.  Upon replacing 
$^{16}$O by $^{18}$O, the peak position shifts down by about 26.8 
cm$^{-1}$.  The energy of the peak for the $^{16}$O sample is about 352 meV.  
Substituting $\hbar\omega_{IR}$ = 352 meV and $J$ = 114 meV into Eq.~\ref{IR} 
yields $\hbar\omega_{ph}$ = 23 meV, close to the values for both 
Sr$_{2}$CuO$_{2}$Cl$_{2}$ and La$_{2}$CuO$_{4}$.  It is interesting to 
note that optical conductivity data of YBa$_{2}$Cu$_{3}$O$_{6.95}$ are 
consistent with very strong coupling to a bosonic mode whose energy is lower 
than 32 meV (Ref.~\cite{Dord}).  If one takes the realistic value of the 
superconducting gap to be 31 meV (Ref.~\cite{Chen}) rather than 25 meV 
used in Ref.~\cite{Dord}, one should obtain the mode energy of about 24 
meV, in quantitative agreement with $\hbar\omega_{ph}$ = 23 meV.  Moreover, the 
coupling strength of this bosonic mode is found to be independent of 
magnetic field, suggesting that this mode is not associated with the 
magnetic resonance mode \cite{Lee}.

The average energy 
of these low-energy phonon modes should not have a significant 
oxygen-isotope shift since the weight of oxygen vibrations for these  modes is less than 30$\%$
\cite{Nozaki}.  We expect that,
upon replacing $^{16}$O by
$^{18}$O, $\hbar\omega_{ph}$ should shift down by $\simeq$0.5 meV.  
Then from the oxygen-isotope shift (3.38 meV) of $\hbar\omega_{IR}$, 
we readily find that $\Delta J/J$ $\simeq$ $-$0.9$\%$.  The magnitude 
of the isotope effect on $J$ for YBa$_{2}$Cu$_{3}$O$_{6}$ is slightly 
larger than that for La$_{2}$CuO$_{4}$ ($\Delta J/J$ $\simeq$ 
$-$0.6$\%$).

Now let's use Eq.~2 to deduce the value of $E_{p}^{O}$ for both
La$_{2}$CuO$_{4}$ and YBa$_{2}$Cu$_{3}$O$_{6}$. The charge-transfer
gaps $\Delta_{pd}$ have been measured for both systems \cite{Cooper}, 
that is,
$\Delta_{pd}$ = 1.81 eV for La$_{2}$CuO$_{4}$ and $\Delta_{pd}$ = 1.60
eV for YBa$_{2}$Cu$_{3}$O$_{6}$. If we take $\hbar\omega_{O}$ =
0.075 eV and substitute the above parameters into Eq.~2, we obtain
$E_{p}^{O}$ = 1.5 eV for La$_{2}$CuO$_{4}$ and $E_{p}^{O}$ = 1.7 eV
for YBa$_{2}$Cu$_{3}$O$_{6}$.  Although the $E_{p}^{O}$ values for the 
two systems are similar, the oxygen-isotope effect on $J$ is 
significantly larger in YBa$_{2}$Cu$_{3}$O$_{6}$ due 
to a smaller $\Delta_{pd}$.  Since 
$E_{p}^{Cu}$ $\neq$ 0, the deduced $E_{p}^{O}$ values should be the 
lower limit of the total polaron binding energy.

Very recently, angle-resolved photoemission spectroscopy (ARPES) data 
of undoped La$_{2}$CuO$_{4}$ have been explained in terms of polaronic
coupling between phonons and charge carriers \cite{Ros}.  From the
width of the phonon side band in the ARPES spectra, the authors find
the polaron binding energy to be about 1.9 eV, in good agreement with
their theoretical calculation based on a shell model \cite{Ros}.  On 
the other
hand, the observed binding energy of the side band should be
consistent with a polaron binding energy of about 1.0 eV 
(Ref.~\cite{Ros}).
This should be the lower limit because the binding energy of the side 
band
decreases rapidly with doping and because the sample may be lightly 
doped
\cite{Ros}.  Therefore, the ARPES data suggest that 1.0 eV $<$ $E_{p}$
$<$ 1.9 eV, which is in quantitative agreement with the value deduced
from the isotope effect on the exchange energy.

The parameter-free estimate of the polaron
binding
energy due to the long-range Fr\"ohlich-type electron-phonon
interaction
has been made for many oxides including cuprates and manganites
\cite{Alex99}. The
polaron binding energy due to the long-range Fr\"ohlich-type 
electron-phonon
interaction is estimated to be about 0.65 eV for La$_{2}$CuO$_{4}$
(Ref.~\cite{Alex99}). The polaron
binding energy due to the $Q_{1}$-type Jahn-Teller distortion is about
1.2 eV for La$_{2}$CuO$_{4}$ (Ref.\cite{Kam}). The total polaron 
binding energy should
be about 1.85 eV, in excellent agreement with the value deduced
from the isotope effect on $J$ and the ARPES data.

If there are very small amounts of charged carriers in these nearly 
undoped compounds, the optical conductivity will show a broad peak at 
$E_{m}$ = 2$\gamma E_{p}$ (Ref.\cite{Alex99}), where 
$\gamma$ is 0.2$-$0.3 for layered cuprates \cite{Alex99}.  The 
$\gamma$ value will be further reduced in the case of $\hbar\omega/t$ 
$<$$<$ 1, where $t$ is the bare hopping integral. There appears to exist the third broad 
peak at 0.7-0.8 eV in the optical conductivity of 
Sr$_{2}$CuO$_{2}$Cl$_{2}$, La$_{2}$CuO$_{4}$, and 
YBa$_{2}$Cu$_{3}$O$_{6}$.  This peak should be caused by the polaronic 
effect because the energy scale for the peak is similar to that 
predicted from the polaron theory assuming $\gamma$ $\sim$ 0.2.  Hole 
doping will reduce the value of $E_{p}$ and thus $E_{m}$ due to 
screening of charged carriers.  Indeed, $E_{m}$ was found to be about 
0.6 eV for La$_{1.98}$Sr$_{0.02}$CuO$_{4}$ and 0.44 eV for 
La$_{1.94}$Sr$_{0.06}$CuO$_{4}$ (Ref.~ \cite{Bi}).

From the inferred polaron binding energy in the parent compounds, we 
can estimate a dimensionless coupling constant $\lambda$ for the 
high-energy phonon modes using $\lambda$ = $E_{p}/zt$, where $z$ is the 
number of the nearest neighbors \cite{Alex2}.  With $E_{p}$ = 2 eV, 
$t$ = 0.4 eV, and $z$ = 4, we find $\lambda$ = 1.25.  The coupling 
constant is not large enough to lead to a structural instability.  It 
has been shown that \cite{Alex92} there is no structural instability 
even at a very large electron-phonon coupling.  This is because when 
small polarons are formed, the phonon frequency renormalization is 
negligible at any carrier density \cite{Alex92}. Moreover, a static long-range 
charge ordering is unlikely to occur in doped systems due to this 
intermediate electron-phonon coupling and the quasi-two-dimensional 
electronic structure.

Now we discuss the isotope effect on the antiferromagnetic ordering
temperature $T_{N}$ in hole-doped La$_{2}$CuO$_{4+y}$ and 
YBa$_{2}$Cu$_{3}$O$_{6+y}$.
It is known that the antiferromagnetic properties of
La$_{2}$CuO$_{4+y}$ can be well understood within
mean-field theory which leads to a $T_{N}$ formula
\cite{Thio}:
\begin{equation}\label{Ie5}
k_{B}T_{N} \sim J' [\xi(T_{N})/a]^{2},
\end{equation}
where $J'$ is the interlayer coupling energy, $\xi (T_{N})$ is the
in-plane AF correlation length at $T_{N}$, which is given by
$\xi (T_{N}) \propto \exp (J/T_{N})$ for undoped compounds with the
maximum $T_{N}$. When $T_{N}$ is
reduced to about 80$\%$ of the maximum $T_{N}$ by doping, a 
mesoscopic phase
separation has taken place so that $\xi (T_{N})$ = $L$
(Ref.~\cite{Cho}), where $L$ is the size of the
antiferromagnetically correlated clusters, and
depends only on the carrier density. In this case, we have $T_{N}
\sim J'(L/a)^{2}$, which is independent of $J$. This can naturally 
explain a
negligible oxygen-isotope effect on $T_{N}$ in La$_{2}$CuO$_{4+y}$ 
with
$T_{N}$ $\simeq$ 250 K (Ref.~\cite{ZhaoAF}).  Because the maximum 
$T_{N}$ is 500 K in
YBa$_{2}$Cu$_{3}$O$_{6+y}$ (Refs.~\cite{Tranq1,Tranq2,Brewer}), the 
negligible oxygen-isotope effect on
$T_{N}$ should be also expected for YBa$_{2}$Cu$_{3}$O$_{6+y}$ with 
$T_{N}$ $\leq$
400 K.

In summary, we have deduced the oxygen-isotope
effect on the in-plane antiferromagnetic exchange energy $J$
for YBa$_{2}$Cu$_{3}$O$_{6}$ from the observed oxygen-isotope effect 
on the
mid-infrared two-magnon absorption peak \cite{Grun}. The
exchange energy $J$ in YBa$_{2}$Cu$_{3}$O$_{6}$ is found to decrease 
by
about 0.9$\%$ upon replacing $^{16}$O by $^{18}$O, which is slightly
larger than that (0.6$\%$) in La$_{2}$CuO$_{4}$. From the isotope
effect, we quantitatively estimate the lower limit of the polaron binding 
energy, which is about 1.7 eV for YBa$_{2}$Cu$_{3}$O$_{6}$ and 1.5 eV 
for La$_{2}$CuO$_{4}$.  The results are in quantitative agreement with 
the recent ARPES data, optical conductivity data, and the 
parameter-free theoretical estimate.  The 
large polaron binding energy in the insulating parent compounds 
suggests that electron-phonon coupling should also be strong in doped 
superconducting cuprates and play an essential role in 
high-temperature superconductivity.  ~\\
~\\
\noindent
{\bf Acknowledgment:} This research is partly supported by a 
Cottrell Science Award from Research Corporation.

~\\
~\\
$^{*}$ gzhao2@calstatela.edu

\end{document}